\documentclass[journal]{IEEEtran}

\pdfpageattr{/Group << /S /Transparency /I true /CS /DeviceRGB >>}

\usepackage[utf8]{inputenc}
\usepackage[T1]{fontenc}
\usepackage{textcomp}
\usepackage{microtype}
\usepackage{calc}
\usepackage{xcolor,comment} 
\usepackage{footnote}

\usepackage[range-phrase=-,per-mode=symbol-or-fraction,range-units=single,list-units=single,detect-all,binary-units]{siunitx}
\usepackage{silence}
\DeclareSIUnit\byte{Byte}
\DeclareSIUnit\decibeli{dBi}
\DeclareSIUnit\decibelm{dBm}
\DeclareSIUnit\megasamplespersecond{Msps}
\DeclareSIUnit\dbm{dBm}
\DeclareSIUnit\ppm{ppm}
\DeclareSIUnit\watthour{Wh}

\usepackage{csquotes}
\usepackage[backend=biber,style=ieee,doi=false,isbn=false,mincitenames=1,maxcitenames=2]{biblatex}
\DeclareFieldFormat{sentencecase}{#1} 
\DeclareFieldFormat{titlecase}{#1} 
\usepackage{xpatch}
\xpatchbibmacro{textcite}{\addspace}{\addnbspace}{}{}
\xpatchbibmacro{Textcite}{\addspace}{\addnbspace}{}{}

\DefineBibliographyStrings{english}{
	andothers = et~al\adddot\addspace
}

\usepackage{amsmath}
\usepackage{amssymb}
\usepackage{amsfonts}
\usepackage{amsthm}

\usepackage{booktabs}
\usepackage{url}

\usepackage[caption=false,font=footnotesize]{subfig}
\usepackage[pdftex]{graphicx}
\DeclareGraphicsExtensions{.pdf,.png,.jpg}
\usepackage{tikz}
\usetikzlibrary{calc}

\usepackage[american]{babel}
\usepackage{hyphenat}

\usepackage{algorithmic}
\usepackage{algorithm}

\usepackage[capitalize,noabbrev]{cleveref}

\crefname{algocf}{Algorithm}{Algorithms}
\Crefname{algocf}{Algorithm}{Algorithms}

\makeatletter
\RequirePackage{xstring}
\RequirePackage{xparse}
\RequirePackage[]{acro}
\RequirePackage{ifthen}
\@ifpackagelater{acro}{2020/04/29}{
\NewDocumentCommand\acrodef{mO{#1}mG{}}{\DeclareAcronym{#1}{short={#2}, long={#3}, #4}}
}{
\@ifpackagelater{acro}{2019/02/17}{
	\NewDocumentCommand\acrodef{mO{#1}mG{}}{\DeclareAcronym{#1}{short={#2}, long={#3}, foreign-plural={}, #4}}
}{
	\NewDocumentCommand\acrodef{mO{#1}mG{}}{\DeclareAcronym{#1}{short={#2}, long={#3}, #4}}
}}
\makeatother
\acrodef{AP}{access point}
\acrodef{AR}{augmented reality}
\acrodef{COTS}{commercially available off-the-shelf}
\acrodef{CPS}{cyber-physical systems}
\acrodef{IoT}{internet of things}
\acrodef{ISP}{internet service provider}
\acrodef{KPI}{key performance indicator}
\acrodef{MEC}{multi-access edge computing}
\acrodef{ML}{machine learning}
\acrodef{M2M}{machine-to-machine}
\acrodef{NFV}{network function virtualization}
\acrodef{QoS}{quality of service}
\acrodef{RAN}{radio access network}
\acrodef{RF}{radio frequency}
\acrodef{SAE}{Society of Automotive Engineers}
\acrodef{VLC}{visible light communication}
\acrodef{VNF}{virtual network function}
\acrodef{V-Edge}{virtual edge computing}
\acrodef{XR}{extended reality}
\acrodef{6DoF}{six degrees of freedom}
\acrodef{VETO}{V-Edge tetrahedron orchestrator}

\usepackage{listings}
\usepackage[textsize=small]{todonotes}

\newcommand{\falko}[1] {\todo[color=green!20,inline]{FD: #1}}

\begin{document}

\title{V-Edge: Virtual Edge Computing to Dynamically Combine Cloud, Edge, and Fog Resources}
\title{V-Edge: Virtual Edge Computing as an Enabler for Novel Microservices and Cooperative Computing}

\author{Falko Dressler, Carla Fabiana Chiasserini, Frank H.P. Fitzek, Holger Karl, Renato Lo Cigno, Antonio Capone, Claudio Casetti, Francesco Malandrino, Vincenzo Mancuso, Florian Klingler, Gianluca Rizzo%
\thanks{This work has been supported in part by the German Research Foundation (DFG) under grant no.\ DR 639/25-1.}
\thanks{F.\ Dressler is with the School of Electrical Engineering and Computer Science, TU Berlin, Berlin, Germany, e-mail: dressler@ccs-labs.org.}%
\thanks{C.\,F.\ Chiasserini is with the Dept.\ of Electronics and Telecommunications, Politecnico di Torino, Torino, Italy, e-mail: carla.chiasserini@polito.it.}%
\thanks{F.\ H.P.\ Fitzek is with the School of Engineering Sciences as well as the excellence cluster CeTI, TU Dresden, Dresden, Germany, e-mail: frank.fitzek@tu-dresden.de.}%
\thanks{H.\ Karl is with the Hasso-Plattner-Institute, Potsdam, Germany, e-mail: holger.karl@uni-paderborn.de.}%
\thanks{R.\ Lo Cigno is with the Dept.\ of Information Engineering, University of Brescia, Brescia, Italy, e-mail: renato.locigno@unibs.it.}%
\thanks{A.\ Capone is with the Dept.\ of Electronics, Information and Bioengineering, Politecnico di Milano, Milano, Italy, e-mail: antonio.capone@polimi.it.}%
\thanks{C.\ Casetti is with the Dept.\ of Control and Computer Engineering, Politecnico di Torino, Torino, Italy, e-mail: claudio.casetti@polito.it.}%
\thanks{F.\ Malandrino is with the Institute of Electronics, Computer and Communication Engineering of the National Research Council of Italy, Torino, Italy, e-mail: francesco.malandrino@ieiit.cnr.it.}%
\thanks{V.\ Mancuso is with IMDEA Networks, Madrid, Spain, e-mail: vincenzo.mancuso@imdea.org.}%
\thanks{F.\ Klingler is with the Dept.\ of Computer Science, Paderborn University, Paderborn, Germany, e-mail: florian.klingler@uni-paderborn.de.}%
\thanks{G.\ Rizzo is with the Research Institute of Information Systems, HES-SO, Valais-Wallis, Switzerland, and with University of Foggia, Italy, e-mail: gianluca.rizzo@hevs.ch.}%
}

\maketitle

\begin{abstract}
As we move from 5G to 6G, edge computing is one of the concepts that needs revisiting.
Its core idea is still intriguing:
instead of sending all data and tasks from an end user's device to the cloud, possibly covering thousands of kilometers and introducing delays that are just owed to limited propagation speed, edge servers deployed in close proximity to the user, e.g., at some 5G gNB, serve as proxy for the cloud.
Yet this promising idea is hampered by the limited availability of such edge servers.
In this paper, we discuss a way forward, namely the \ac{V-Edge} concept.
\Ac{V-Edge} bridges the gap between cloud, edge, and fog by virtualizing all available resources including the end users' devices and making these resources widely available using well-defined interfaces.
\Ac{V-Edge} also acts as an enabler for novel microservices as well as cooperative computing solutions.
We introduce the general \ac{V-Edge} architecture and we characterize some of the key research challenges to overcome, in order to enable wide-spread and even more powerful edge services.
\end{abstract}
\acresetall

%

\section{Introduction}
\label{sec:introduction}

New-generation mobile networks are envisioned to provide the computational, memory, and storage resources needed to run services required by diverse third parties (referred to as vertical industries or verticals).
Each service is associated with specific requirements, quantified as \acp{KPI}.
To this end, networks will require a high degree of flexibility and fully automated operations, with a drastically reduced service deployment time.
Essential components to achieve these goals are softwarization of both networking and services using \ac{NFV}~\cite{wang2016network,agarwal2019vnf,alam2020survey} and the ability to store and process data close to the end user leveraging the so-called edge computing paradigm~\cite{mach2017mobile}.
Note that edge computing is not just about classic \ac{MEC}~\cite{jiang2019toward,mach2017mobile}; rather, the network edge has become the convergence point of data processing, caching, and communication~\cite{wang2017survey}, which makes service provisioning at the edge one of the key challenges in future networks.

Network virtualization, greatly supported by the current 5G/6G standardization and research beyond it, pushes \ac{NFV} to merge with the concept of microservices to improve practicality, universality, and automation.
%
Service ubiquity, resilience, and low latency are emerging as the ultimate goals -- following up the recent work in the context of the Tactile Internet~\cite{dressler2019cooperative,fitzek2021tactile}.
To achieve these goals, networks are progressively integrating \ac{ML}~\cite{letaief2019roadmap} in two main ways.
First, an increasing number of user applications include \ac{ML} models for a smarter application behavior, higher ability to adapt to  user's preferences, and more effective interaction between users and machines.
Second, ML-based approaches have become common in automatic network management, resource orchestration, as well as predicting a wide range of parameters (e.g., wireless  channel properties, users' behavior, service demand).

\begin{figure}
	\centering
	\includegraphics[width=\columnwidth]{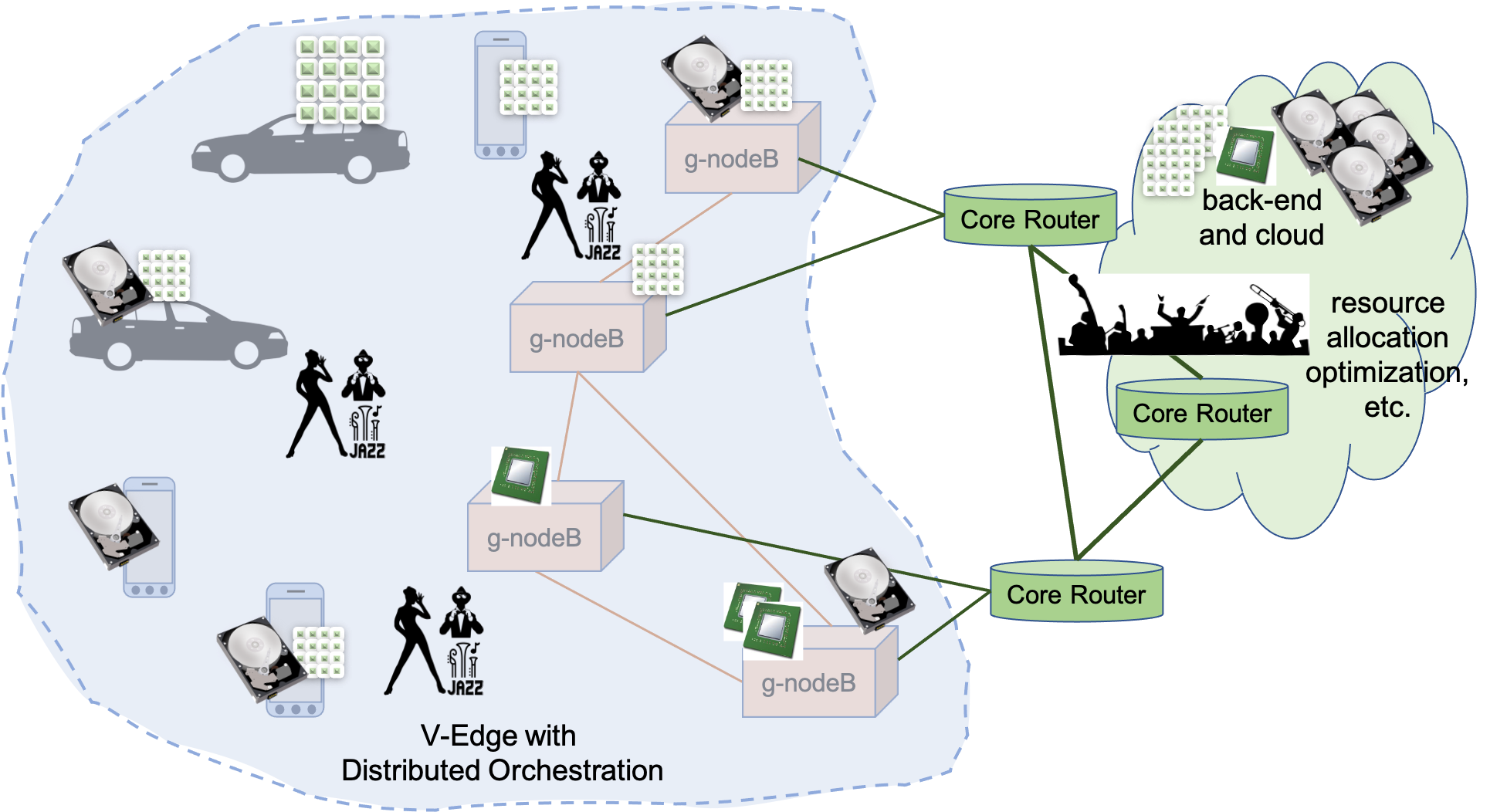}
	\caption{The V-Edge concept: Abstraction of the physical resources of the cloud back-end, the 5G core and \ac{RAN}, as well as users and machines. All components provide (and may use) resources for edge computing and they may participate in a local, distributed orchestration on short time-scale, while global optimization and other non-real-time operations can be performed in the cloud back-end.}
	\label{fig:concept}
	\vspace{-.5em}
\end{figure}

\begin{table*}
\centering
\caption{Typical nodes participating in V-Edge clusters}
\label{tab:nodes}
\begin{tabular}{lllll}
\toprule
\multicolumn{5}{l}{\textbf{Stationary / infrastructure-based systems}} \\
Type & CPU & Storage & Network & Time available \\
\midrule
ISP-operated MEC server & high-performance multi-core & $\SIrange{1}{100}{\tera\byte}$ & $\SIrange{0.1}{10}{\giga\bit\per\second}$ & years \\
privately operated MEC server & multi-core & $\SIrange{1}{10}{\tera\byte}$ & $\SIrange{0.1}{1}{\giga\bit\per\second}$ & weeks \\
Wi-Fi APs & single core & $\SIrange{0.1}{100}{\giga\byte}$ & $\SIrange{0.1}{1}{\giga\bit\per\second}$ & months\\
\midrule
\multicolumn{5}{l}{\textbf{Mobile / opportunistic systems}} \\
Type & CPU & Storage & Network & Time available \\
\midrule
Moving car & low-end multi-core & $\SIrange{0.1}{1}{\tera\byte}$ & $\SIrange{0.1}{1}{\giga\bit\per\second}$ & $\SIrange{1}{5}{\minute}$ \\
Parked car & low-end multi-core & $\SIrange{0.1}{1}{\tera\byte}$ & $\SIrange{0.1}{1}{\giga\bit\per\second}$ & $\SIrange{0.5}{24}{\hour}$ \\
Fully autonomous shuttle & high-performance multi-core & $\SIrange{1}{10}{\tera\byte}$ & $\SIrange{0.1}{1}{\giga\bit\per\second}$ & $\SIrange{1}{30}{\minute}$ \\
User with smartphone & single- to multi-core & $\SIrange{1}{100}{\giga\byte}$ & $\SIrange{0.1}{1}{\giga\bit\per\second}$ & $\SIrange{5}{15}{\minute}$ \\
\bottomrule
\end{tabular}
\end{table*}

Given these emerging trends, \emph{the network edge is turning into an enabler} between the cloud and a fully-distributed \ac{M2M} network, hosting virtualized network functions and user applications, to meet both service providers' and users' needs. 
%

This work introduces the \emph{\acf{V-Edge}} concept.
It takes advantage of the flexibility offered by network softwarization and \ac{NFV} to integrate in an opportunistic and dynamic manner the highly heterogeneous set of resources available locally at the edge (e.g., computing, storage, and communication resources), while guaranteeing seamless and QoS-aware service provisioning to users in a variety of verticals.
Further, it jointly uses such resources for any virtualized function of a user application.

A schematic representation of the \ac{V-Edge} concept is depicted in \cref{fig:concept}.
Compared to 5G and traditional edge computing, the system comprises dynamic resources: CPUs, connectivity, and storage capacity come and go as users do, carrying the corresponding devices.
Thus, we have to move from allocating \emph{static} resources to dynamic users and applications to allocating  resources that are dynamic as well.
An example is the integration of cars not only as service users, but also as service providers such as explored in the vehicular micro-cloud concept~\cite{dressler2019virtual}.
\ac{V-Edge} goes well beyond initial activities towards distributed computing and data storage, realizing a full and harmonic integration between infrastructure-based communication networks and mobile edge systems at the resource level, as well as between user applications and network functions at the service layer.

In \ac{V-Edge}, part of the orchestration of resources and tasks needs to be done at the edge on rather short time scales to cope with resource volatility and dynamics.
The back-end cloud, instead, can be used for global optimization on longer time scales.
Following current \ac{ML} approaches to 5G and edge computing~\cite{letaief2019roadmap}, \ac{V-Edge} will also be inherently learning-based, supporting both user applications and network functions.
\ac{V-Edge} can implement privacy-preserving, distributed approaches such as federated learning~\cite{wang2019adaptive}) and effectively transfer trained model where and as needed.

Our main contributions can be summarized as follows:
\begin{itemize}
\item we characterize the need to go beyond classic \ac{MEC} for higher scalability, resilience, and flexibility;
\item we introduce the conceptual architecture of \ac{V-Edge} making consequent use of virtualization to deal with the high degree of dynamics in the network; and
\item we discuss relevant research questions to be solved to make \ac{V-Edge} reality.
\end{itemize}

%

\section{The \acs{V-Edge} Ecosystem}
\label{sec:eco-system}

Before outlining the virtual edge computing architecture, we introduce the underlying basic components of the \ac{V-Edge} eco-system, including the major services it can support.

\paragraph{Users}
As in conventional systems, users still contribute to the traffic demand while using edge-based applications.
In \ac{V-Edge}, users may have a dual representation in the system as edge users but also as resource providers.

\paragraph{Resources}
The resources to satisfy the user demand, network-wise and application-wise, are now provided by an increasing variety of devices ranging from the cloud to ISP-operated edge servers, and even to community-operated edge devices and to smaller IoT systems.
\ac{V-Edge} thus goes well beyond classic \ac{MEC}, by dropping the differentiation between cloud and edge and fog, and opportunistically recruiting local, already existing -- yet possibly unused -- resources.
In \ac{V-Edge}, even small ``fog'' devices are conceptually turned into ``edge servers'' to provide functions to third parties.
A list of typical \ac{V-Edge} computational, storage, and networking resources is provided in \Cref{tab:nodes}, which also indicates the average time each kind of edge node will be available in a given location.
This results in dynamic scenarios -- classic \ac{MEC} assumes dynamics in terms of users and their tasks coming and going.
Now, also the available edge computing resources come and go in a very dynamic way, and  they can be seen as constituting a \emph{virtualized edge server} with time-varying resource availability.

\paragraph{Services and Functions} Network services, and often user applications, need to be deployed within the \ac{V-Edge}.
The classes of user applications that can benefit most from a virtual edge implementation are:
\begin{itemize}
	\item services with tight latency constraints or whose support with dedicated static infrastructure would have brought too high CAPEX, e.g., cooperative (automated) driving and UAV control, in need of capillary local edge support even out of cities;
	\item services that may exhibit bursts of demand of computing tasks, e.g., due to ``flash crowds'';
	\item non-latency-constrained applications such as computing tasks for situation awareness, distributed video processing, and event detection;
	\item \ac{IoT} applications like monitoring tasks, where local data have to be pre-processed for immediate use or the transferring of large amounts of data to the cloud would require too much bandwidth;
	\item \ac{AR}, and in general \ac{XR}, applications, as well as all \ac{6DoF} immersive technology  that require both low latency and large data rate;
	\item \ac{ML} applications making use of the ML as a service~\cite{bacciu2017need} concept, which has indeed emerged as a new paradigm, whereby trained or pre-trained models are provided for making decisions of different type and in different contexts.
\end{itemize}

\paragraph{Orchestrator}
To complete the above functions, resource and service orchestration is needed.
Resource orchestration can be both reactive (which may sometimes be too late) or  proactive, so that resources, and the functions mapped thereon, can naturally follow demand in space and time.
We remark that the orchestration itself becomes one of the tasks to be distributed and executed within the virtual edge, similarly in this respect to user applications.

\begin{figure}
	\centering
	\includegraphics[width=\columnwidth]{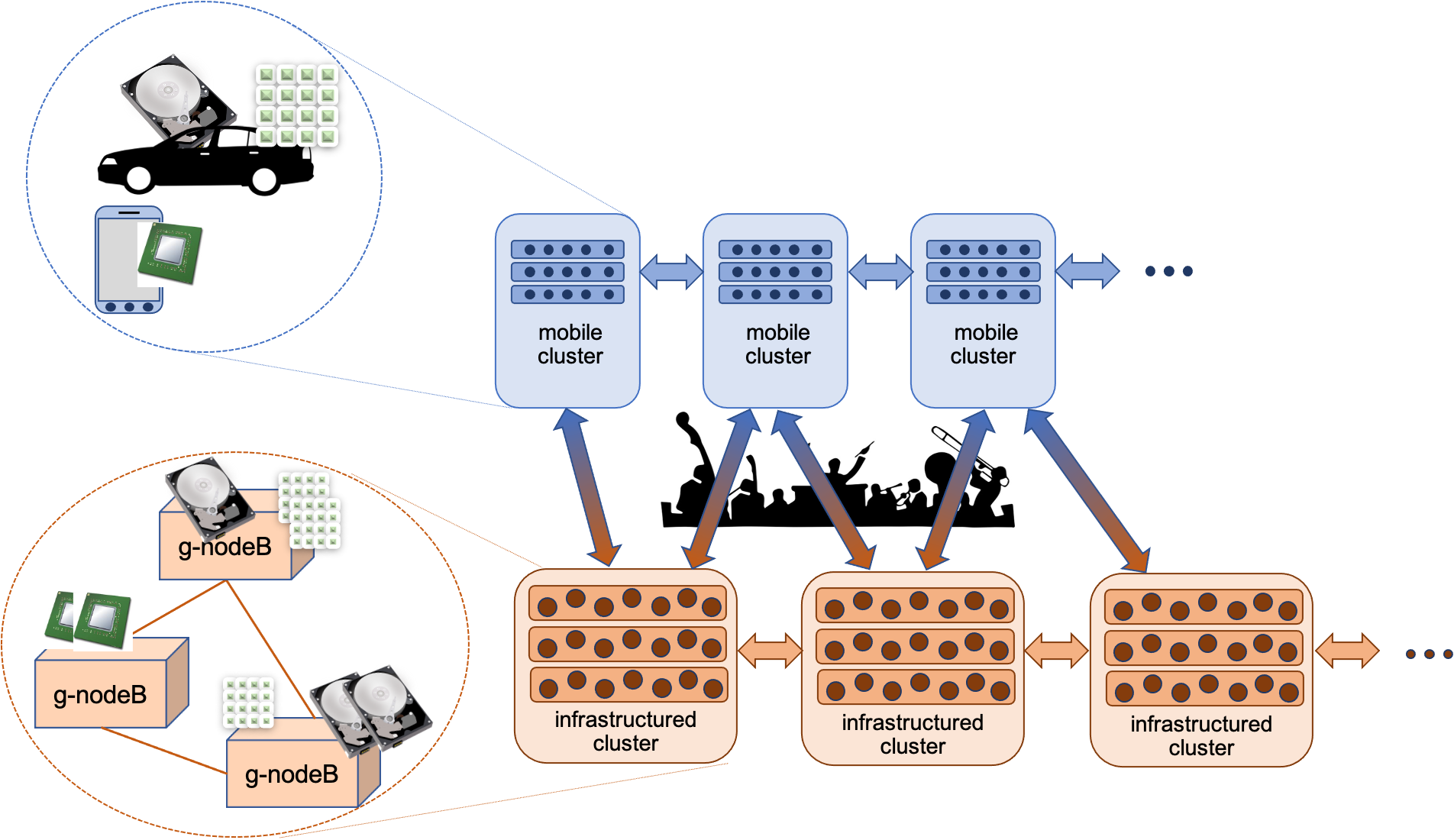}
	\caption{V-Edge architecture: Logical resources from mobile users and infrastructure-based systems (e.g., edge servers co-located with a gNB), are aggregated into clusters. Multiple clusters are appropriately coordinated and microservices can migrate from one cluster to another to optimize the service location.
	Resource management is done by an orchestrator, which may interwork with others, controlling neighboring clusters, to migrate services.}
	\label{fig:architecture}
\end{figure}

The orchestrator (see\ \cref{fig:architecture}) has to observe and monitor nodes and their computing and communication resources, and schedule them for current functions and microservices.
Machine learning will help to make such decision with little and often impaired information about the available edge components.
From an architectural perspective, the orchestrator can be centralized at a (physical) edge server (or even in the cloud, with the risk of additional problems due to the inherent communication delay), or decentralized through hierarchically-coordinated clusters of nodes participating in the \ac{V-Edge}.
In realistic deployments, a partially distributed solution may be preferred for better resilience and responsiveness of the overall system.

\paragraph{Architecture}
The architecture of a \ac{V-Edge} system enables the interaction between the above described basic components, as illustrated in \Cref{fig:concept,fig:architecture}.
A key feature of the \ac{V-Edge} architecture is that users are virtually clustered so as to provide resources qualitatively equivalent to the ones provided by the infrastructure.
This cluster-based organization is meant to facilitate and optimize resource management while providing resiliency and flexibility, like done, e.g., in the context of vehicular micro-clouds~\cite{dressler2019virtual}.
Services and network functions can be instantiated in a cluster and then migrated to another one dynamically, under the coordination of the orchestrator, and as a consequence of the learning process that underpins its operations. 
\Cref{fig:architecture} zooms into the architecture outlining interconnected mobile and infrastructure clusters that are orchestrated together.
The distributed nature of all resources additionally requires novel concepts and interfaces for distributed orchestration and for the cooperation between orchestrators, and even between multiple such clusters, edge components, and the back-end cloud servers.

%

\section{Key Technologies and Research Challenges}
\label{sec:key-technologies}

Existing work on edge computing has predominantly focused on resource allocation on edge servers that may experience dynamic load, but whose deployment is static or only changes on a long time scale.
As mentioned, \ac{V-Edge} goes well beyond and lifts this limitation by allowing also servers to be mobile, thus, computational, storage, and communication resources appear and leave at any time.
In this section, we identify the most relevant key technologies that can make \ac{V-Edge} a reality.

%

\subsection{Performance Aspects}
\label{sec:perf-aspects}

Similarly to non-virtual edge clouds, a \ac{V-Edge} system needs to support \acp{KPI} such as high throughput, low latency, low service rejection rate, high utilization, short provisioning times, high dependability, and easy management, as well as to maximize the number of satisfied users and expected revenue compared to capital or operational expenditure (CAPEX, OPEX).
There exists, however, a differentiating factor between \ac{V-Edge} and non-virtual edge clouds: the node \emph{churn rate} in the underlying network and the  evolution of the network topology in space and time due to devices joining and  leaving the \ac{V-Edge}.
Here, churn rate is not a performance metric, rather, a system characteristic with a twofold impact.
On the one hand, it may lead to a degradation in the \ac{V-Edge} \acp{KPI}, which  could be characterized as the \emph{price of virtuality};
on the other hand, the contribution of mobile devices to the \ac{V-Edge} allows for significant CAPEX and OPEX savings.

While this is a fair perspective from an end user's or investor's perspective, it can fall short when comparing different \ac{V-Edge} realizations against each other.
First, more fine-grained metrics would be needed in this case to characterize the performance of services as well as management and orchestration systems (e.g., packet latency vs.\ service initiation time, or traffic throughput vs.\ number of service deployments per second).
Second, suitable metrics should be selected to highlight the existing trade-offs in performance.
A typical example is the overhead introduced by state synchronization to ward off service interruptions, compared against the degradation in the users' quality of experience caused by those same service interruptions.

\begin{figure}
  \centering
   \includegraphics[width=1\columnwidth]{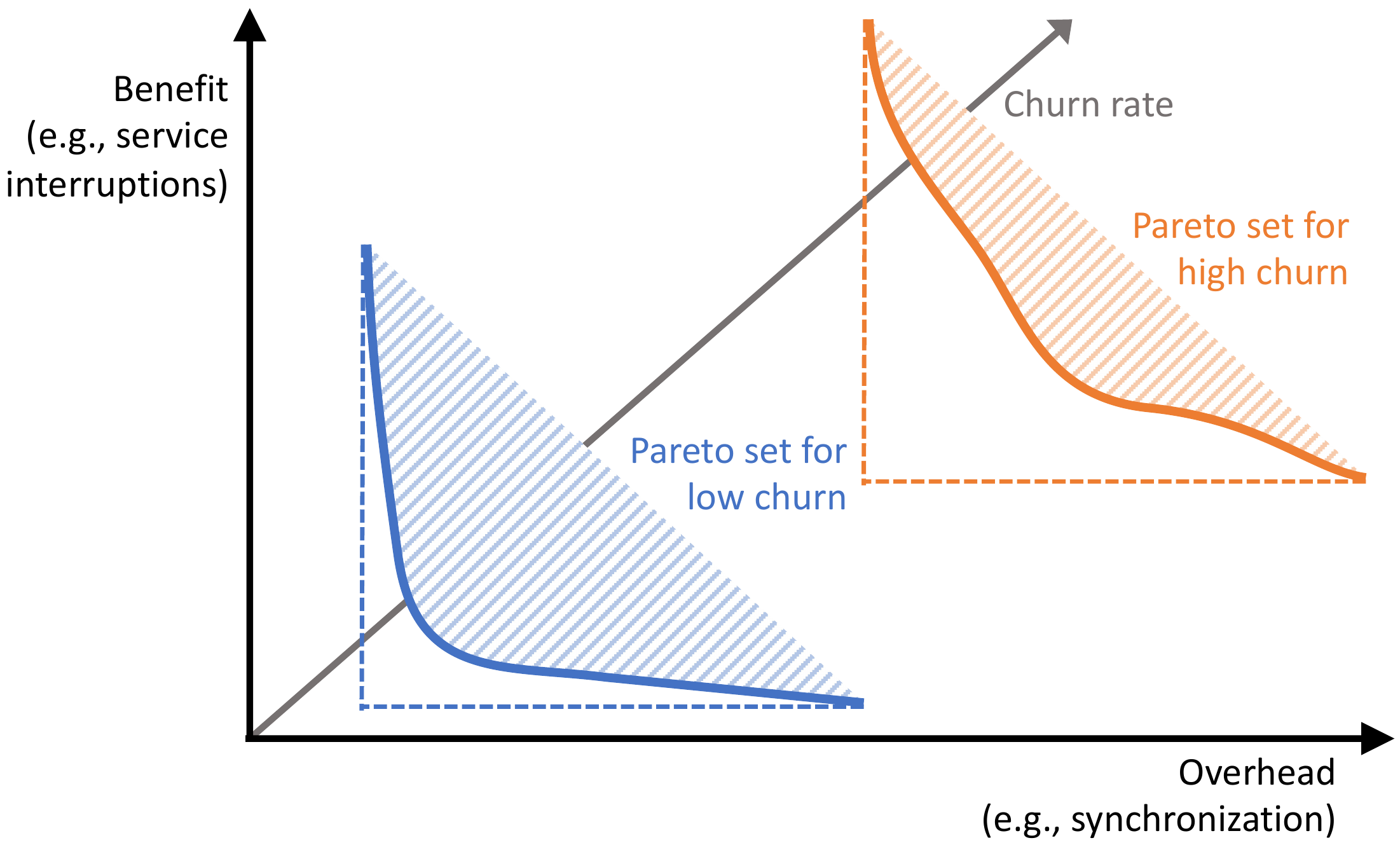}
  \caption{Services \acp{KPI} in the context of the \ac{V-Edge} concept.}
  \label{fig:kpis_services}
\end{figure}

We argue that different trade-offs can be achieved depending upon how a \ac{V-Edge} is configured, obtaining a Pareto front of optimal trade-offs.
Further, as exemplified in \cref{fig:kpis_services}, different Pareto fronts may emerge for different churn rates and devices' connectivity patterns.
Such complex dependencies and entanglement between diverse aspects of the \ac{V-Edge} cannot be captured through the existing performance metrics; they  rather call for a novel, complex notion of ``figure of merit''.

%

\subsection{Orchestration of Microservices}
\label{sec:micr-netw-funct}

Network softwarization is taking over the data, control, and management planes as well as different protocol layers.
Examples of data plane virtualization include virtual routers and user applications, while a relevant control plane example is the O-RAN control of the radio interface.
As mentioned in \cref{sec:eco-system}, \acp{VNF} stemming from such softwarization can be seen as (components of) microservices, which need to be properly and jointly orchestrated whenever they compete for the same physical resources.
Further, depending on their type and logic, microservices can be executed in different execution environments with varying trade-offs in terms of capabilities and performance.

Thus, an orchestrator for a \ac{V-Edge} system needs to provide the same functionality as any of the orchestrators proposed for an ordinary edge infrastructure.
Namely, it has to map \acp{VNF} from microservices to the available resources,  taking into account not only their requirements but also the computing and communication capabilities of the device on which they are mapped and the  performance impact of the services that leverage such microservice instances.
This is, however, not the only issue a \ac{V-Edge} orchestrator faces.
Indeed, it  has to cope  with  the network and node churn: quickly changing network conditions and node availability.
A \ac{V-Edge} orchestrator has to be aware of this churn as well as of the services' ability to deal or not to deal with it (e.g., stateless vs.\ stateful services) and their temporal and spatial availability requirements -- aspects that are exacerbated in \ac{V-Edge} with respect to conventional scenarios.
This fact invalidates any conventional, long-term approach and demands for a more agile, adaptive solution.


We address this challenge by leveraging \ac{ML} techniques, conceiving a multi-faceted framework that can effectively deal with the multitude of necessary observations and actions.
Specifically, the proposed \ac{V-Edge} orchestration framework includes:
\begin{enumerate}
\item a network model, partially based on explicit information (e.g., battery or computing capacity of a device) and partially learned information (e.g., movement patterns and sojourn time), to account for  individual devices' capabilities and behavior;
\item a service model, partially provided by the \ac{VNF} graph composing the service and the VNFs' specifications, partially learned (e.g., how disruption-tolerant is a service, how does a disruption affect the users' quality of experience);
we underline that some information that could be provided by the service developer might actually need to be learned in practice and that a  continuous update of the service understanding is necessary;
\item the orchestrator as such: learning scaling, placement, routing, migration,  and other actions based on the network and service models;
\item an Auto-ML component: since the above three models need to be continuously trained in the field and since properly parameterizing training is hard, an Auto-ML component is necessary to tune training hyperparameters.
\end{enumerate}

\begin{figure}
  \centering
  \includegraphics[width=0.9\columnwidth]{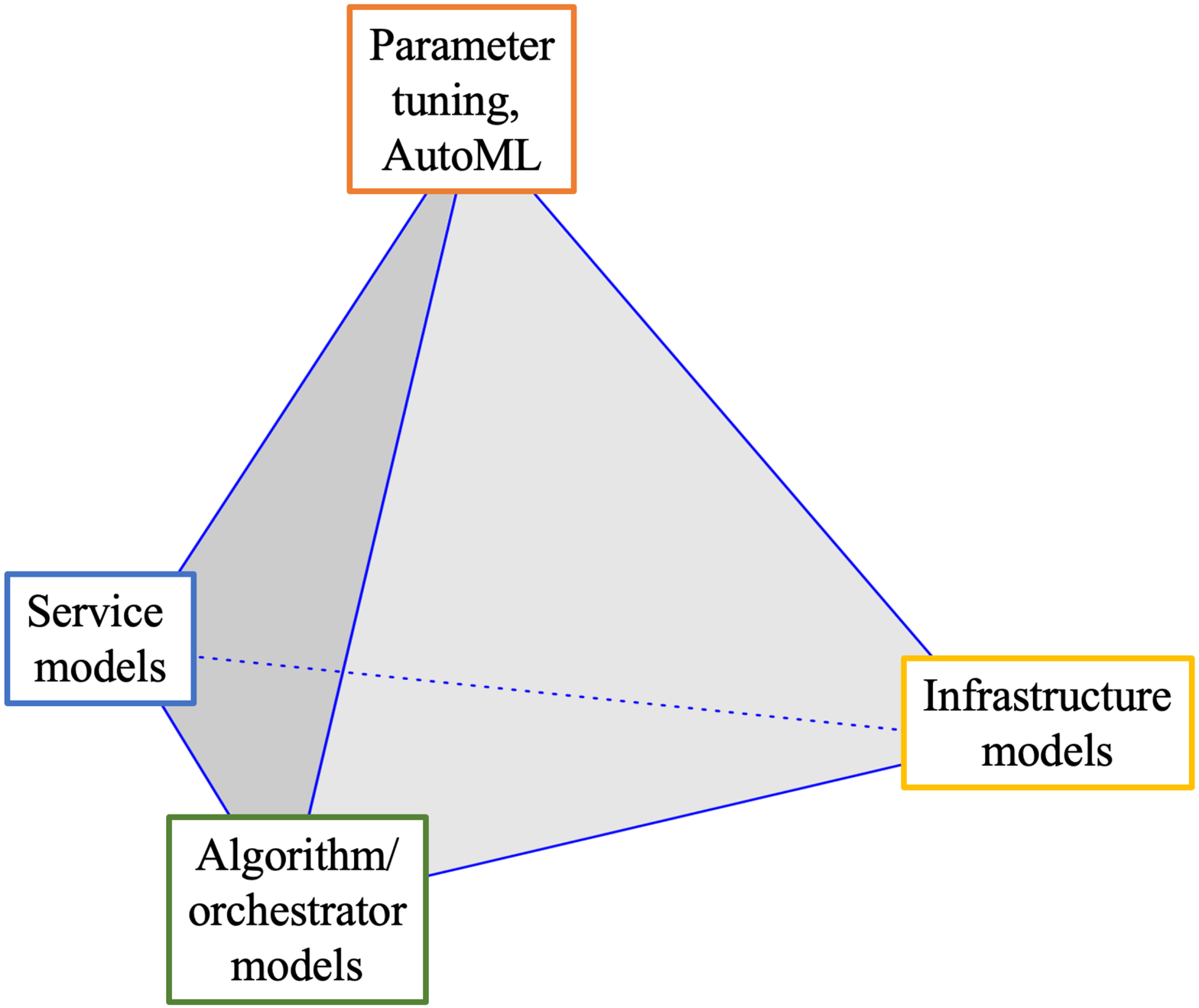}
  \caption{The \acf{VETO} concept.}
  \label{fig:veto}
\end{figure}

The  four components of the framework are connected in a tetrahedron as depicted in \cref{fig:veto}, as they all depend on each other; we dub such a framework \ac{VETO}.
\ac{VETO} provides a functional separation of  a learning-based orchestrator.
Some important challenges, however, remain to be addressed for a detailed framework design.
In particular, it is critical to: (i) learn correlation between network and service models, e.g., between user spatial distribution and service demand dynamics, (ii) identify the hyperparameters to be learned by the Auto-ML component, (iii) define the time scale over which the different components should operate, (iv) understand with which granularity instances of VETO should be deployed to deal with different geographical areas to make the system scalable.

%

\subsection{Cooperative Computing}
\label{sec:distr--coop}

Cloud computing has become extremely popular due to its flexible (cost) structures and dynamic resource allocation; overall, it has been a door opener for many novel services.
Edge computing and service placement in close proximity of the user enabled a new kind of services particularly focusing on low latency and often referred to in the context of the Tactile Internet~\cite{dressler2019cooperative,fitzek2021tactile}.

\begin{figure}
	\centering
	\includegraphics[width=\columnwidth]{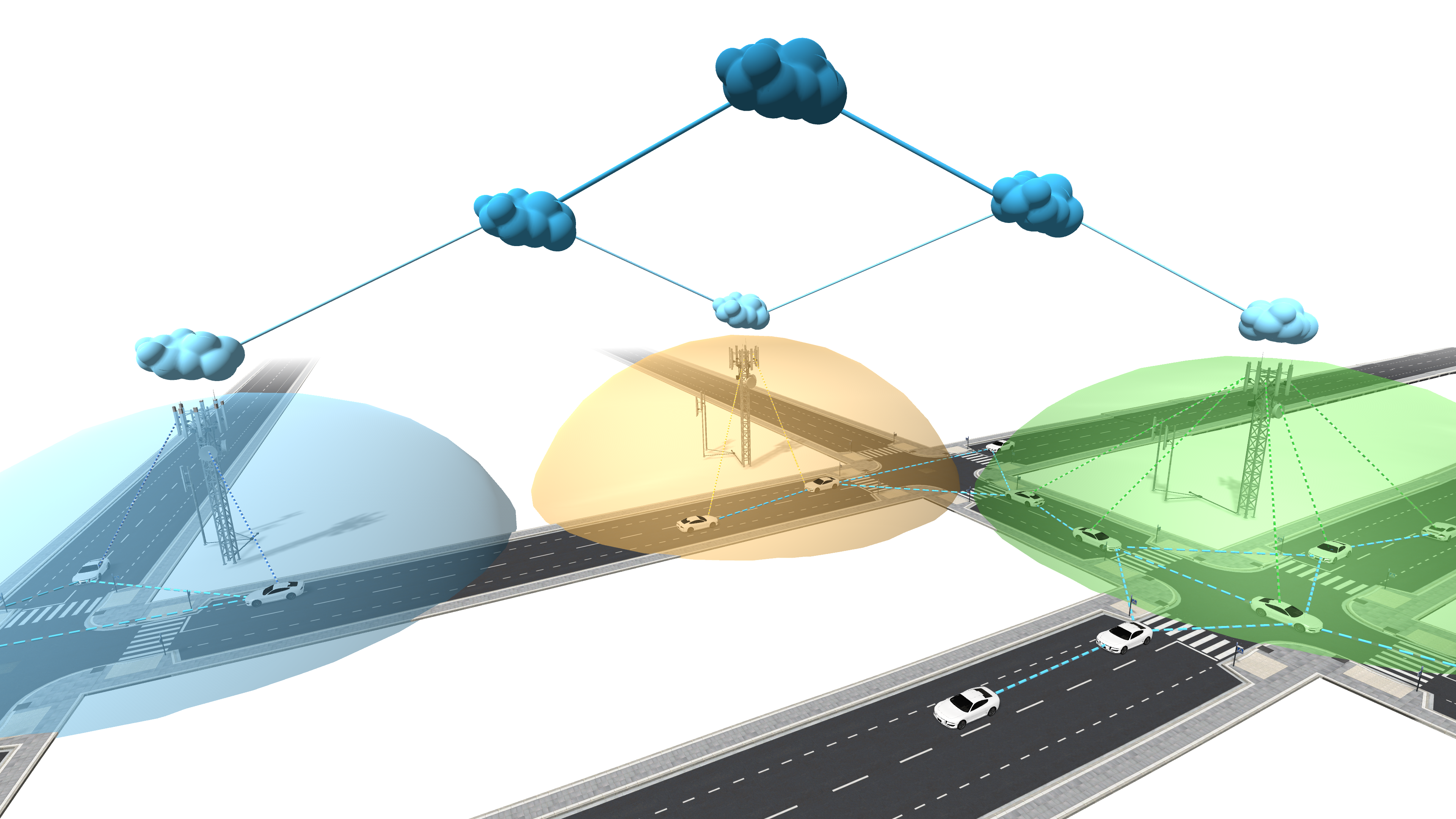}
	\caption{Placement of mobile edge computing for connected vehicles.}
	\label{fig:clouds3Dcars}
\end{figure}

However, latency is not only caused by propagation delays.
Other factors such as computing delay also play a dominant role so that the optimal placement of a function is not always at the edge~\cite{keller2017response}.
\cref{fig:clouds3Dcars} illustrates the problem in  the application scenario of connected vehicles, highlighting different placement options.
The \ac{V-Edge} architecture indeed allows for vertical and horizontal placement and distribution of edge infrastructure.

A vital research field is resilience in such agile environments.
Virtualization of all edge computing components introduces the risk of service starvation: tasks currently offloaded to virtual edge components may not be completed, or not be completed in time.
In communications and storage, replication and redundancy is often used to overcome such problems, and, 
even though not efficient in terms of resource utilization, this can greatly increase resilience of the system. 
Recent advances in the field of coding, in particular network coding and coded caching, help achieving efficiency, resilience, and latency at the same time.
Unfortunately, these ideas do not immediately apply to computing.
Replication has to deal with erroneous feedback information and the state of the cooperative computing instances may diverge.
Microservices running on cooperative machines improve resilience, but not efficiency.

\emph{Coded computing} may provide solutions for cooperative computing~\cite{taik-kim2020coded}.
In order to enable cooperative computing in \ac{V-Edge}, the following phases are to be considered.
\begin{itemize}
\item \emph{Phase 0} is about hierarchical mobile edge clouds as described already for vehicular micro clouds~\cite{dressler2019virtual}.
Edge nodes are interconnected in form of clusters in the \ac{V-Edge} architecture to distribute tasks among themselves using the backend data center as a fallback solution.
All relevant meta data describing tasks and associating end users need to be synchronized among all participating nodes in a hierarchical manner.

\item \emph{Phase 1} uses a coding-based approach.
Similar to network coding for storage and communication, also computing tasks can be coded to avoid outages if physical nodes leave the virtual edge~\cite{taik-kim2020coded}.
Coded computing is normally used between neighboring nodes such as mobile robots or cars, but it can be extended to cooperation among multiple edge clouds, adding resilience and performance.

\item In a final \emph{Phase 2}, such coding-based distributed computing will be inherently integrated with new approaches to resource management.
Current resource management solutions focus on communication, sometimes also incorporating computational resources, but still in a simplistic way.
\end{itemize}

For cooperative computing, either explicit communication between participating nodes in the \ac{V-Edge} or mediation by hierarchies up to the backend data center, or implicit communication by means of inference is needed.
\Cref{fig:clouds3Dcars} illustrates the concept of localized edge clouds interconnected in a hierarchical manner to perform such cooperative computing.
Research questions range from the identification of required data, to finding paths the data has to travel along, to data fusion, and compressed sensing.
Cooperative computing, of course, has to rely on distributed learning concepts.
Here, federated learning will play a dominant role because of its capability to train distributedly and then to merge the resulting models in a privacy preserving manner~\cite{sattler2020robust,wang2019adaptive}.
The grand challenge here is sparsification of models in order to reduce the overhead resulting from model distribution.

\section{Discussions and Conclusion}
\label{sec:conclus}

Revisiting the motivation for our \acf{V-Edge} approach, in the following we discuss the benefits of this novel architecture and what cannot be done with traditional cloud, \ac{MEC}, and possibly fog computing.
From a paradigmatic point of view, there are many reasons to make use of \ac{V-Edge}, though some fundamental problems need to be addressed before implementation.

From a \textit{policy perspective}, the \ac{V-Edge} concept addresses many problems that have hampered (mobile) global communications in the past decades.
As discussed above, \ac{V-Edge} requires necessarily open solutions like O-RAN, or similar ones, at different architectural levels.
Openness in telecommunications and computing has proven to be one of the key enablers for innovation and economic growth; thus, the \ac{V-Edge} vision naturally becomes the melting pot for novel services, solutions, start-ups, and technological evolution.
This consideration alone should be enough for all actors, and standardization bodies in particular, to embrace \ac{V-Edge} and mold future business based on this equitable architecture.

\textit{Scalability} is a more technical reason to foster \ac{V-Edge}.
5G/6G architectures, together with computation (think about GPUs) and local access (WiFi~6 and the upcoming WiFi~7), have shown that only extreme distribution and densification of resources can meet the increasing requirements on communications and services.
\ac{V-Edge} is bringing this evidence from subliminal awareness to architectural design, highlighting and formalizing the interdependence between communications, computing, management (resource allocation and scheduling), and service \acp{KPI}.
So doing, \ac{V-Edge} clarifies the technical challenges that need to be addressed for success, first of all in the realm of \ac{ML}, acknowledging that traditional models cannot be applied to a system whose evolution is not predictable \emph{a priori}. It also allows for a level of adaptability and flexibility that empowers the autonomous evolution of functions, services, and management models through autonomous learning and self-development.

\textit{Efficient resource utilization} is going hand in hand with scalability to make advanced services more accessible and, thus, affordable by a wider sector of the society.
If, on the one hand, a distributed architecture is the only solution to scalability, on the other hand, it is well known that uncoordinated distribution makes an inefficient use of the resources, from storage and computing power, down to communications and energy.
\ac{V-Edge} proposes an advanced, \ac{ML}-oriented orchestration that enables the efficient and dynamic use of resources, especially leveraging those that go unused for a large part of the time.
An example for all: processing power on autonomous vehicles when they are parked.
The safety requirements of \ac{SAE} Level-5 autonomous driving require a processing power (CPU and GPU) that is comparable to several nodes of high performance computing systems, and this extreme capacity is right there, at the edge, but with traditional architectures it is impossible to tap it.

Finally, \textit{security and privacy} need to be considered, which goes well beyond the scope of this paper.
Theoretical computer science indicates that distributed systems are in general safer, more secure, and most of all naturally following the implementation of ``privacy by design'' principles.
\ac{V-Edge} clearly matches this indication, with its extreme distribution and the orchestration of resources coming from different actors and entities.
However, we are also well aware that \textit{practical} systems often fail to meet theoretical results, in particular in the case of security where the complexity of the analysis of distributed systems may lead to design failures, with severe consequences. 
This is a further topic for research and design towards the \ac{V-Edge} realization.

%

\printbibliography

%

\begin{IEEEbiographynophoto}{Falko Dressler} (F'17)
is full professor and Chair for Telecommunication Networks at TU Berlin.
His research objectives include adaptive wireless networking (sub-6GHz, mmWave, visible light, molecular communication) and wireless-based sensing.
\end{IEEEbiographynophoto}

\begin{IEEEbiographynophoto}{Carla~Fabiana Chiasserini} (F'18) 
is full Professor  with Politecnico di Torino and a Research Associate with CNR-IEIIT. 
Her research interests include architectures, protocols, and performance analysis of wireless networks.
\end{IEEEbiographynophoto}

\begin{IEEEbiographynophoto}{Frank H. P. Fitzek}
is leading the Deutsche Telekom Chair of Communication Networks. Current research directions are Tactile Internet, Quantum Communication, and Post Shannon with a general interest in SDN, NFV, ICN, network coding, cooperative networks, and robotics. The chair is leading the excellence cluster Tactile Internet with Human-in-the-Loop (CeTI).
\end{IEEEbiographynophoto}

\begin{IEEEbiographynophoto}{Holger Karl}
currently holds the chair of IT-Technology and Systems at the Hasso-Plattner-Institute, Potsdam. Before, from 2004 to 2021, he was chair of Computer Networks at Paderborn University. His interests are network softwarization, data centres, and mobile/wireless networks.
\end{IEEEbiographynophoto}

\begin{IEEEbiographynophoto}{Renato Lo Cigno} (SM)
is Full Professor at the University of Brescia, Italy, where he leads the Advanced Networking Systems group. His interests are in wireless and vehicular networks, performance evaluation, and novel communication paradigms and architectures. 
\end{IEEEbiographynophoto}

\begin{IEEEbiographynophoto}{Antonio Capone} (F'19) 
is full Professor at Politecnico di Milano and Dean of the School of Industrial and Information Engineering. His research areas include network softwarization and data plane programmability, resource optimization and planning of wireless networks. 
\end{IEEEbiographynophoto}

\begin{IEEEbiographynophoto}{Claudio Casetti}
is a full professor with Politecnico di Torino, Turin, Italy. 
His research interests include vehicular networks, 5G networks, IoT and has published over 200 papers on these topics.
He is Senior Editor of IEEE Vehicular Technology Magazine.
\end{IEEEbiographynophoto}

\begin{IEEEbiographynophoto}{Francesco Malandrino} (SM'19)
is a researcher at the National Research Council of Italy (CNR-IEIIT). His research interests include the architecture and management of wireless, cellular, and  vehicular networks.
\end{IEEEbiographynophoto}

\begin{IEEEbiographynophoto}{Vincenzo Mancuso} (SM'21)
is Research Associate Professor at IMDEA Networks Institute, Spain. His interests include  analysis, design, and experimental evaluation of opportunistic architectures for intelligent wireless networks and edge computing.
\end{IEEEbiographynophoto}

\begin{IEEEbiographynophoto}{Florian Klingler}
received his Ph.D. degree (Dr. rer. nat., summa cum laude) in Computer Science from Paderborn University, Germany, in 2018.
His research focuses on protocols for adaptive wireless networks in highly dynamic scenarios.
\end{IEEEbiographynophoto}

\begin{IEEEbiographynophoto}{Gianluca Rizzo}
is Associate Professor at the University of Foggia, Italy, and Senior Research Associate at HES SO Valais, Switzerland.
His research interests are in performance evaluation of Computer Networks.
\end{IEEEbiographynophoto}

\vfill

\end{document}